\documentstyle[12pt]{article}
\newcommand{\CR}{\nonumber \\}
\newcommand{\pa}{\partial}
\newcommand{\A}{\alpha}
\newcommand{\B}{\beta}
\newcommand{\D}{\delta}
\newcommand{\G}{\gamma}

\newcommand{\cN}{{\cal N}}
\newcommand{\cL}{{\cal L}}
\newcommand{\cG}{{\cal G}}
\newcommand{\cT}{{\cal T}}
\newcommand{\vp}{\varphi}
\renewcommand{\thefootnote}{\fnsymbol{footnote}}

\begin{document}
\begin{titlepage}
\begin{flushright}
hep-th/9811002 \\
YITP-98-74 \\
October, 1998
\end{flushright}
\vspace{0.5cm}
\begin{center}
{\Large \bf 
Extended Superconformal Algebras on $AdS_{3}$
}
\lineskip .75em
\vskip2.5cm
{\large Katsushi Ito}
\vskip 1.5em
{\large\it Yukawa Institute
for Theoretical Physics \\  Kyoto University, Kyoto 606-8502, Japan}  
\vskip 3.5em
\end{center}
\vskip3cm
\begin{abstract}
We study a supersymmetric extension of the Virasoro algebra 
on the boundary  of the anti-de Sitter space-time $AdS_{3}$. 
Using the free field realization of the currents, 
we show that the world-sheet affine Lie superalgebras 
$osp(1|2)^{(1)}$, $sl(1|2)^{(1)}$ and $sl(2|2)^{(1)}$ provide the
boundary $\cN=1,2$ and $4$ extended superconformal algebras, respectively.

\end{abstract}
\end{titlepage}
\baselineskip=0.7cm
\newpage
\renewcommand{\thefootnote}{\arabic{footnote}}
The duality between the type IIB string theory on $AdS_{3}\times S^{3}\times 
M^{4}$, where $M^{4}=K3$ or $T^{4}$, and two-dimensional
$\cN=4$ superconformal field theory on a symmetric product of $M^{4}$
\cite{StMa,ads,Gi}, is one of
interesting  examples of the AdS/CFT correspondence\cite{Ma}. 
Conformal symmetry on the boundary of $AdS_{3}$, firstly introduced by
Brown and Henneaux \cite{BrHe}, is realized as the chiral algebra of 
the boundary conformal field theory associated with the $SL(2,R)$ 
Chern-Simons theory\cite{CS}.
On the other hand, Giveon et al. \cite{Gi} constructed the boundary Virasoro 
algebra from the string theory on $AdS_{3}$. 
The generators of the algebra are identified as the global charges
associated with the world-sheet $sl(2,R)$ current algebra, which is expressed 
in terms of the free fields\cite{Wa}.
They also constructed (a part of) the $\cN=4$ superconformal algebra 
{}from the superstrings on $AdS_{3}\times S^{3}\times T^{4}$.
Their boundary algebra, however, is not manifestly supersymmetric
since the supercurrents are introduced by the bosonization of the 
world-sheet fermions and the algebra
is defined up to the picture changing operator\cite{FMS}.

In the present paper, we study the boundary extended superconformal
algebra realized in a manifestly supersymmetric way.
By replacing the world-sheet affine Lie algebra $sl(2,R)$ to an affine
Lie superalgebra which includes the subalgebra $sl(2,R)$ and employing 
the free field realization of the currents, we will obtain
the extended superconformal algebra which acts on the boundary 
$AdS_{3}$ superspace. 
In particular, we will show that the world-sheet affine Lie superalgebras
$osp(1|2)^{(1)}$, $sl(1|2)^{(1)}$ and $sl(2|2)^{(1)}$ provide the
boundary $\cN=1,2$ and $4$ extended superconformal algebras, respectively.

We begin with reviewing the boundary Virasoro algebra associated with 
$AdS_{3}$. 
The anti-de Sitter space $AdS_{3}$ is the hypersurface 
$-U^{2}-V^{2}+X^{2}+Y^{2}=-\ell^{2}$ embedded in the flat space
$R^{2,2}$.
In the coordinates $(\rho,\tau,\phi)$ defined by $U=\ell\cosh\rho\sin\tau$,
$V=\ell\cosh\rho\cos\tau$, $X=\ell \sinh \rho \cos\phi$,
$Y=\ell\cosh\rho \sin\phi$, the metric is given by
\begin{equation}
{ds^{2}\over \ell^{2}}=
-\cosh^{2}\rho d\tau^2+\sinh^{2}\rho d\phi^{2}+d\rho^{2}.
\label{eq:met}
\end{equation}
In the $AdS_{3}$ space, there is $SL(2,R)_{L}\times SL(2,R)_{R}$
symmetry.
The generators of $SL(2,R)_{L}$ read \cite{StMa}
\begin{equation}
L_{0}= i \pa_{u}, \quad 
L_{\pm 1}= i e^{\pm i u} \left( \coth 2 \rho \pa_{u}-
{1\over \sinh 2\rho} \pa_{v}\mp {i\over 2}\pa_{\rho}\right),
\end{equation}
where $u=\tau+\phi$, $v=\tau-\phi$. 
$SL(2,R)_{R}$ is generated by $\bar{L}_{0,\pm1}$, which are obtained
by $u\leftrightarrow v$.
Define the coordinates $(\tilde{\phi},\G,\bar{\G})$ by
\begin{eqnarray}
\tilde{\phi}&=& \ell \cosh \rho -i \tau, \CR
\G&=& \tanh \rho e^{i u}, \quad \bar{\G}=\tanh \rho e^{i v}.
\end{eqnarray}
The metric (\ref{eq:met}) becomes
\begin{equation}
ds^{2}=\ell^{2} (d\tilde{\phi}^{2}+e^{2\tilde{\phi}}d\G d\bar{\G}). 
\end{equation}
The $SL(2,R)_{L}$ generators are 
\begin{eqnarray}
L_{0}&=& {1 \over2}\pa\tilde{\phi}-\G\pa_{\G}, \CR
L_{-1}&=& -\pa_{\G}, \CR
L_{1}&=& \G \pa\tilde{\phi}-\G^{2}\pa_{\G}+
\ell^{2}e^{-2\tilde{\phi}}\pa_{\bar{\G}}.
\end{eqnarray}
Thus in the limit $\tilde{\phi}\rightarrow \infty$, the derivative 
with respect to
$\bar{\G}$ in $L_{1}$, decouples and the $SL(2,R)$
generators become holomorphic with respect to  $\G$. 
\footnote{ Here we consider the Euclidean version of the theory, 
which is obtained by replacing $\tau$ by $-i\tau_{E}$.}
These currents are regarded as the zero-mode part
of the Wakimoto realization \cite{Wa} of the affine
Lie algebra $sl(2,R)$ at level $k$:
\begin{eqnarray}
H(z)&=& -i\sqrt{2}\A_{+}\pa\vp +2\G\B, \CR
J_{-}(z)&=& \B, \CR
J_{+}(z)&=&i\sqrt{2}\A_{+}\pa\vp\G-\G^{2}\B-k\pa\G,
\end{eqnarray}
where $\A_{+}=\sqrt{-k+2}$.
$(\B(z),\G(z))$ is the bosonic ghost system with conformal weights
$(1,0)$ and has the operator product expansion (OPE)
$\B(z)\G(w)=1/(z-w)+\cdots$. $\vp(z)$ is a free boson with the OPE
$\vp (z)\vp (w) =-\log (z-w)+\cdots $. 
Then 
\begin{equation}
L_{0}= -{1\over2} \oint {dz \over 2\pi i} H(z), \quad
L_{\pm 1}= \pm\oint {dz \over 2\pi i} J_{\pm}(z),
\end{equation}
generate $sl(2,R)$. For integer $n$, it is shown in \cite{Gi}
 that the generators
\begin{equation}
L_{n}= \oint {dz \over 2\pi i} \cL_{n}(z),
\label{eq:genvir}
\end{equation}
where
\begin{equation}
\cL_{n}(z)= -{1-n^{2}\over 2} H \G^{n}
-{n(n-1)\over2} J_{-}\G^{n+1}
+{n(n+1)\over2} J_{+}\G^{n-1},
\label{eq:cln}
\end{equation}
satisfy the Virasoro algebra
\begin{equation}
\mbox{[} L_{m}, L_{n} \mbox{]}= (m-n) L_{m+n}+{c\over 12} (m^{3}-m)
\delta_{m+n,0}, 
\label{eq:vir}
\end{equation}
with central charge $c=6kp$.
Here the contour integral in (\ref{eq:genvir}) is taken around $z=0$
and $p$ is defined by
\begin{equation}
p\equiv \oint {d z\over 2\pi i} {\pa\G\over \G}.
\end{equation}

We now consider an $\cN$ supersymmetric extension of this Virasoro
algebra (\ref{eq:vir}), which is realized in the superspace
$(\G, \xi_{1}, \cdots, \xi_{\cN})$. 
Here $\xi_{1}, \cdots, \xi_{\cN}$ denote the Grassmann odd 
coordinates. 
The supercharges are realized as the conserved charges of the 
anti-commuting currents with spin one on the world-sheet, 
whose zero modes should obey the Lie superalgebras
$osp(1|2)$, $osp(2|2)=sl(1|2)$ and $sl(2|2)$ for $\cN=1,2,4$ 
supersymmetries respectively.
Hence we expect that conformal field theory whose chiral algebra is 
the affine Lie superalgebra with a subalgebra $sl(2,R)$,
provides a natural supersymmetric generalization of the boundary 
conformal symmetry.
In the following we will construct some boundary superconformal
algebras explicitly.
The affine Lie superalgebra ${\bf g}$ is generated by the bosonic currents
$J_{\A}(z)$ ($\A\in\Delta_{0}$), $H^{i}(z)$ ($i=1,\cdots, r$)
and fermionic currents $j_{\B}(z)$ ($\B\in\Delta_{1}$).
Here $H^{i}(z)$ are the Cartan currents and 
$\Delta_{0}$ ($\Delta_{1}$)  the set of even (odd) roots.

The free field representation \cite{LS} of an affine Lie superalgebra 
with rank $r$ is obtained by $r$ free bosons $\phi^{i}(z)$ ($i=1,\cdots, r$), 
pairs of bosonic ghosts $(\B_{\A}(z), \G_{\A}(z))$ with conformal
weights $(1,0)$ for positive
even roots $\A$ and pairs of fermionic ghosts 
$(\eta_{\A}(z),\xi_{\A}(z))$ with conformal weights $(1,0)$ for 
positive odd roots $\A$. The OPEs
for these fields are
$\phi^{i}(z)\phi^{j}(w)=-\delta_{i j}\log (z-w)+\cdots $,
$\B_{\A}(z)\G_{\A'}(w)=\delta_{\A,\A'}/(z-w)+\cdots $, 
$\eta_{\A}(z)\xi_{\A'}(w)=\delta_{\A,\A'}/(z-w)+\cdots $.

Firstly we consider the $\cN=1$ superconformal symmetry.
The Lie superalgebra $osp(1|2)$ contains odd roots $\pm \A_{1}$ and 
even roots $\pm 2\A_{1}$ with $\A_{1}^{2}=1$.
The free field realization of affine Lie superalgebra $osp(1|2)^{(1)}$
at level $-k$ is 
\begin{eqnarray}
j_{\A_{1}}(z)&=& (-2k+2) \pa \xi+2 i \A_{+} \xi \pa\vp +\G\eta-2\xi\G\B, \CR
J_{2\A_{1}}(z)&=& {-k\over2}\pa\G
-(-k+2)\xi\pa\xi+i\A_{+}\G\pa\vp-\G^{2}\B
-\G\xi\eta,
\CR
j_{-\A_{1}}(z)&=& \eta-2\xi\B, \quad 
J_{-2\A_{1}}(z)= \B,\CR
\A_{1}\cdot H(z)&=& -i \A_{+}\pa \vp +2\G\B+\xi\eta,
\end{eqnarray}
where 
$(\eta,\xi)=(\eta_{\A_{1}},\xi_{\A_{1}})$,
$(\B,\G)=(\B_{2\A_{1}},\G_{2\A_{1}})$
and $\A_{+}=\sqrt{-k+3}$.
This algebra contains the $sl(2,R)$ current algebra 
which is generated by $H(z)=\A_{1}\cdot H(z),J_{\pm}(z)=J_{\pm 2\A_{1}}(z)$.
Then the generators  $\cL_{n}(z)$ defined by (\ref{eq:cln})
satisfy the Virasoro algebra (\ref{eq:vir}).
Concerning the  supercurrent, let us introduce the conserved charges 
\begin{equation}
G_{\pm {1\over2}}={1\over \sqrt{2}}\oint {d z \over 2\pi i} j_{\pm \A_{1}}(z).
\end{equation}
{}From the OPE for the currents, it is shown that $L_{n}$ ($n=0,\pm 1$) 
and $G_{\pm 1/2}$ satisfy the Lie superalgebra $osp(1|2)$.
Higher modes of the supercurrent are obtained by extracting the simple
pole term in the OPE between $\cL_{n}(z)$ and $j_{-\A_{1}}(z)$, which 
is denoted by $\cG_{n-1/2}(z)$. 
The result is 
\begin{equation}
\cG_{n-{1\over2}}(z)={n\over\sqrt{2}} j_{\A_{1}} \G^{n-1}
-{n-1\over\sqrt{2}} j_{-\A_{1}}\G^{n}
-\sqrt{2}n (n-1)\left( H \G^{n-1}\xi+J_{+}\G^{n-2}\xi -J_{-}\G^{n}\xi\right).
\label{eq:sup1}
\end{equation}
We define the generators $G_{n-1/2}$ by 
\begin{equation}
G_{n-{1\over2}}=\oint {dz\over 2\pi i} \cG_{n-{1\over2}}(z).
\label{eq:scrr}
\end{equation}
It is shown that the generators $L_{n}$ and $G_{r}$ obey the 
$\cN=1$ superconformal algebra in the NS sector:
\begin{eqnarray}
\mbox{[} L_{m}, L_{n} \mbox{]}&=& (m-n) L_{m+n}+{c\over 12} (m^{3}-m)
\delta_{m+n,0}, \CR
\mbox{[} L_{n}, G_{r} \mbox{]}&=& \left( {n\over 2}-r\right) G_{n+r}, \CR
\{ G_{r}, G_{s} \}&=& 2 L_{r+s}+{c\over 3} (r^{2}-{1\over 4})
\delta_{r+s,0},
\quad  (n,m \in {\bf Z}, \quad r,s \in {\bf Z}+{1\over 2})
\label{eq:scr1}
\end{eqnarray}
with the central charge $c=6kp$. 
For example, the third eq. in (\ref{eq:scr1}) follows from OPEs
\begin{eqnarray}
j_{\A_{1}}(z)j_{\A_{1}}(w)&=& {4 J_{2\A_{1}}(w)\over z-w}+\cdots,
\quad 
j_{-\A_{1}}(z)j_{-\A_{1}}(w)= {-4 J_{-2\A_{1}}(w)\over z-w}+\cdots, \CR
j_{\A_{1}}(z)j_{-\A_{1}}(w)&=& {2 k\over (z-w)^{2}}+
{-2 \A_{1}\cdot H(w)\over z-w}+\cdots, \CR
j_{\A_{1}}(z)\G(w)&=& {-2\G\xi(w)\over z-w}+\cdots, 
\quad j_{\A_{1}}(z)\xi(w)= {\G(w)\over z-w}+\cdots,  \CR
j_{-\A_{1}}(z)\G(w)&=& {-2\xi(w)\over z-w}+\cdots, 
\quad j_{-\A_{1}}(z)\xi(w)= {1\over z-w}+\cdots,
\end{eqnarray}
and $(H\G^{n-1}+J_{+}\G^{n-2}-J_{-}\G^{n})\xi=-k/2 \G^{n-2}\pa\G\xi$. 
Note that the third term in (\ref{eq:sup1}) is necessary to obtain 
the correct superconformal algebra, although in the bosonic case
the combination $-H\G-J_{+}+J_{-}\G^{2}$ becomes the total derivative
and may be regarded as the longitudinal part.
Let us take the zero mode part of the supercurrent and 
the Virasoro generator.
Regarding $\B$  and $\eta$ as the derivatives ${\pa\over \pa\G}$ and 
${\pa\over \pa\xi}$ respectively, we find
\begin{eqnarray}
G_{n-1/2}&\sim & {1\over \sqrt{2}} (\G^{n}{\pa\over \pa\xi}- 2
\G^{n}\xi {\pa\over \pa\G}), \CR
L_{n}&\sim& -\G^{n+1}{\pa\over
\pa\G}-{n+1\over2}\G^{n}\xi{\pa\over\pa\xi}. 
\end{eqnarray}
Thus, in this classical limit, the ${\cal N}=1$ superconformal algebra
is  realized on the superspace with coordinates $(\G,\xi)$.

Now we generalize this result to the case of $\cN=2,4$ extended 
superconformal algebras by considering
affine Lie superalgebras $sl(1|2)^{(1)}$, $sl(2|2)^{(1)}$, respectively.
The Lie superalgebra $sl(N|M)$ has even roots 
$\pm(e_{i}-e_{j})$ ($1\leq i<j\leq N$), 
$\pm(\delta_{a}-\delta_{b})$ ($1\leq a<b\leq M$)
and odd roots
$\pm (e_{i}-\delta_{a})$ ($i=1,\cdots, N$, $a=1,\cdots, M$).
Here $e_{i}$ ($\delta_{a}$) are the orthonormal basis with positive
(negative) metric $e_{i}\cdot e_{j}=\delta_{ij}$
($\delta_{a}\cdot\delta_{b}=-\delta_{ab}$).
We take the simple roots as $\A_{i}=e_{i}-e_{i+1}$, ($i=1,\cdots,
N-1$), 
$\A_{N}=e_{N}-\delta_{1}$,
$\A_{N+a}=\delta_{a}-\delta_{a+1}$ ($a=1,\cdots, M-1$). 
The OPEs for the currents of the affine Lie
superalgebra $sl(N|M)^{(1)}$ at level $k$ are given by
\begin{eqnarray}
& & J_{e_i-e_j}(z)J_{e_k-e_l}(w)=
           {k\D_{j,k}\D_{i,l}\over (z-w)^{2}}\CR
& & +
 {\D_{j,k}(1-\delta_{i,l})J_{e_i-e_l}(w)-\D_{i,l}(1-\delta_{k,i})J_{e_k-e_i}(w)
+\delta_{j,k}\delta_{il}(e_{i}-e_{j})\cdot H(w) \over z-w}+\cdots, \CR
& & J_{\delta_a-\delta_b}(z)J_{\delta_c-\delta_d}(w)=
           {-k\D_{b,c}\D_{a,d}\over (z-w)^{2}} \CR
&&+ 
{\D_{d,a}(1-\delta_{c,b})J_{\delta_c-\delta_b}(w)
-\D_{b,c}(1-\delta_{a,d})J_{\delta_a-\delta_d}(w)
-\delta_{b,c}\delta_{a,d} (\delta_{a}-\delta_{b})\cdot H(w) \over z-w}
    +\cdots, \CR
&& j_{e_{i}-\delta_{a}}(z)j_{\delta_{b}-e_j}(w)
={-k\D_{i,j}\D_{a,b}\over (z-w)^{2}} \CR
&& -{\D_{a,b}(1-\delta_{i,j})J_{e_i-e_j}(w)
     +\D_{i,j}(1-\delta_{a,b})J_{\delta_b-\delta_a}(w)
     -\delta_{i,j}\delta_{a,b} (\delta_{a}-e_{i})\cdot H(w)\over
z-w}+\cdots,
\CR
&& J_{e_i-e_j}(z)j_{e_k-\delta_a)}(w)
={\D_{j,k}j_{e_i-\delta_a}(w)\over z-w}+\cdots,\  \
J_{e_i-e_j}(z)j_{\D_a-\D_k}(w)={-\D_{i,k}j_{\D_a-\D_j}(w)\over z-w}+\cdots,\CR
&& J_{\delta_a-\delta_b}(z)j_{e_k-\delta_c}(w)
={\D_{a,c}j_{e_k-\delta_b}(w)\over z-w}+\cdots,\ \ 
J_{\delta_a-\delta_b}(z)j_{\delta_c-e_k}(w)=
{-\D_{b,c}j_{\delta_a-e_k}(w)\over z-w}+\cdots,
\CR
&& H^{i}(z)H^{j}(w)={k\delta^{ij}\over (z-w)^{2}}+\cdots, \CR
&& H^{i}(z)J_{\A}(w)={\A^{i}J_{\A}(w)\over z-w}+\cdots, \quad
H^{i}(z)j_{\A}(w)={\A^{i}j_{\A}(w)\over z-w}+\cdots . 
\end{eqnarray}
For $M=2$, $H(z)=\A_{N+1}\cdot H(z)$,
$J_{\pm}(z)=J_{\pm\A_{N+1}}(z)$
become the currents of the $sl(2,R)$ subalgebra.
Then it will be shown that (\ref{eq:cln}) for $\G=\G_{\A_{N+1}}$ obeys 
the Virasoro algebra (\ref{eq:vir}). The other even subalgebra
$u(N)^{(1)}$ ($sl(2)^{(1)}$ for $N=2$) turns out to be an affine Lie 
algebra symmetry on the boundary.

Let us consider the $\cN=2$ ($sl(1|2)$) case. 
The free field realization of the affine Lie superalgebra $sl(1|2)^{(1)}$ at 
level $k$ is given by
\begin{eqnarray}
j_{-\A_{1}}(z)&=&\eta_{\A_{1}}, \quad
J_{-\A_{2}}(z)=\B_{\A_{2}}-\xi_{\A_{1}}\eta_{\A_{1}+\A_{2}}, \quad 
j_{-\A_{1}-\A_{2}}(z)= \eta_{\A_{1}+\A_{2}}, \CR
j_{\A_{1}}(z)&=& k\pa\xi_{\A_{1}}+i\A_{+}\A_{1}\cdot\pa\vp\xi_{\A_{1}}
-(\xi_{\A_{1}+\A_{2}}+\xi_{\A_{1}}\G_{\A_{2}})\B_{\A_{2}}
-\xi_{\A_{1}}\xi_{\A_{1}+\A_{2}}\eta_{\A_{1}+\A_{2}}, \CR
J_{\A_{2}}(z)&=& -(k+1)\pa\G_{\A_{2}}-i\A_{+}\A_{2}\cdot\pa\vp \G_{\A_{2}}
-\G_{\A_{2}}^{2}\B_{\A_{2}}-\xi_{\A_{1}+\A_{2}}\eta_{\A_{1}}, \CR
j_{\A_{1}+\A_{2}}(z)&=& k\pa\xi_{\A_{1}+\A_{2}}+
(k+1) (\pa\G_{\A_{2}}) \xi_{\A_{1}}
+i \A_{+}(\A_{1}+\A_{2})\cdot\pa\vp
\xi_{\A_{1}+\A_{2}}
 \CR
& & +i \A_{+}\A_{2}\cdot\pa\vp \G_{\A_{2}}\xi_{\A_{1}}
+\xi_{\A_{1}}\xi_{\A_{1}+\A_{2}}\eta_{\A_{1}}
+\xi_{\A_{1}+\A_{2}}\G_{\A_{2}}\B_{\A_{2}}
+\xi_{\A_{1}}\G_{\A_{2}}^{2}\B_{\A_{2}}, \CR
H^{i}(z)&=& -i\A_{+}\pa\vp^{i}+\A_{1}^{i}\xi_{\A_{1}}\eta_{\A_{1}}
+\A_{2}^{i}\G_{\A_{2}}\B_{\A_{2}}
 +(\A_{1}+\A_{2})^{i}\xi_{\A_{1}+\A_{2}}\eta_{\A_{1}+\A_{2}},
\end{eqnarray}
where  $\A_{+}=\sqrt{k-1}$.
Let us introduce the operator
\begin{equation}
\cL_{n}(z)={1-n^{2}\over 2} \A_{2}\cdot H\G^{n}
-{n(n-1)\over2} J_{-\A_{2}}\G^{n+1}
+{n(n+1)\over2} J_{\A_{2}}\G^{n-1},
\end{equation}
which corresponds to the Virasoro generator.
Here $\G=\G_{\A_{2}}$. 
As in the case of ${\cal N}=1$, from the OPEs between $\cL_{n}(z)$ and 
$j_{\pm \A_{1}}(w)$, one may construct the operators corresponding
to the supercurrents:
\begin{eqnarray}
\cG_{n-1/2}(z)&=& -(n-1) j_{-\A_{1}-\A_{2}}\G^{n}+n
j_{-\A_{1}}\G^{n-1}, \CR
\bar{\cG}_{n-1/2}(z)&=& -(n-1) j_{\A_{1}}\G^{n}-n
j_{\A_{1}+\A_{2}}\G^{n-1} \CR
& &  -
n(n-1) \left( -\A_{2}\cdot H \G^{n-1}\xi_{\A_{1}+\A_{2}}
+J_{-\A_{2}}\G^{n}\xi_{\A_{1}+\A_{2}}-J_{\A_{2}}\G^{n-2}\xi_{\A_{1}+\A_{2}}
\right).
\end{eqnarray}
By extracting the simple pole term in the OPE between
$\cG_{n-1/2}(z)$ and $\bar{\cG}_{1/2}(w)$, one may read off the
operator $\cT_{n}(z)$, which corresponds to the $U(1)$ current. 
The result is 
\begin{equation}
\cT_{n}(z)= -(\A_{1}+2 \A_{2})\cdot H \G^{n}
-n \tilde{\xi}_{\A_{1}+\A_{2}}\tilde{\eta}_{\A_{1}}\G^{n-1},
\end{equation}
where $\tilde{\xi}_{\A_{1}+\A_{2}}=
\xi_{\A_{1}+\A_{2}}+\G_{\A_{2}}\xi_{\A_{1}}$, 
$\tilde{\eta}_{\A_{1}}= \eta_{\A_{1}}-\G_{\A_{2}}\eta_{\A_{1}+\A_{2}}$.
After introducing these operators, we can show that the generators
\begin{eqnarray}
L_{n}&=&\oint {d z\over2\pi i} \cL_{n}(z) , \quad 
T_{n}=\oint {d z\over2\pi i} \cT_{n}(z), \quad (n\in{\bf Z})
\CR
G_{r}&=&\oint {dz\over 2\pi i} \cG_{r}(z), \quad 
\bar{G}_{r}=\oint {dz\over 2\pi i} \bar{\cG}_{r}(z),
\quad (r\in {\bf Z}+1/2)
\end{eqnarray}
satisfy the $\cN=2$ superconformal algebra with central charge $c=6kp$:
\begin{eqnarray}
\mbox{[} L_{m}, L_{n} \mbox{]}&=& (m-n) L_{m+n}+{c\over 12} (m^{3}-m)
\delta_{m+n,0}, \CR
\mbox{[} L_{m}, G_{r} \mbox{]}&=& \left( {m\over2}-r\right) G_{m+r},
\quad 
\mbox{[} L_{m}, \bar{G}_{r} \mbox{]}= \left( {m\over2}-r\right)
\bar{G}_{m+r}, \CR
\mbox{[} L_{m},T_{n} \mbox{]}&=& -n T_{m+n}, \CR
\mbox{[} T_{m}, G_{r}\mbox{]}&=& G_{m+r}, \quad 
\mbox{[} T_{m}, \bar{G}_{r}\mbox{]}= \bar{G}_{m+r}, \CR
\mbox{[} T_{m}, T_{n} \mbox{]}&=& {c\over3} m \delta_{m+n,0}, \CR
\{ G_{r}, G_{s} \}&=& \{ \bar{G}_{r}, \bar{G}_{s} \}=0, \CR
\{ G_{r}, \bar{G}_{s} \}&=& L_{r+s}+{1\over2} (r-s) T_{r+s}+{c\over6}
(r^{2}-{1\over4}) \delta_{r+s,0}, 
\end{eqnarray}
where  $n,m\in{\bf Z}$ and $r,s\in {\bf Z}+1/2$.

Similar construction can be done for $\cN=4$ ($sl(2|2)$) case. 
In terms of free fields, the currents of the affine Lie superalgebra
$sl(2|2)^{(1)}$ at level $k$ are given by
\begin{eqnarray}
J_{-\A_{1}}(z)&=& \B_{\A_{1}}, \quad
j_{-\A_{2}}(z)= \eta_{\A_{2}}-\G_{\A_{1}}\eta_{\A_{1}+\A_{2}}, \CR
J_{-\A_{3}}(z)&=& \B_{\A_{3}}-\xi_{\A_{2}}\eta_{\A_{2}+\A_{3}}
-\xi_{\A_{1}+\A_{2}} \eta_{\A_{1}+\A_{2}+\A_{3}}, \quad
j_{-\A_{1}-\A_{2}}(z)= \eta_{\A_{1}+\A_{2}}, \CR 
j_{-\A_{2}-\A_{3}}(z)&=& \eta_{\A_{2}+\A_{3}}
-\G_{\A_{3}}\eta_{\A_{1}+\A_{2}+\A_{3}}, \quad 
j_{-\A_{1}-\A_{2}-\A_{3}}(z)= \eta_{\A_{1}+\A_{2}+\A_{3}}, \CR
J_{\A_{1}}(z)&=& k\pa\G_{\A_{1}}+i \A_{+}
\A_{1}\cdot\pa\vp\G_{\A_{1}}
-\G_{\A_{1}}^{2}\B_{\A_{1}}
+\G_{\A_{1}}\xi_{\A_{2}}\eta_{\A_{2}}
-\G_{\A_{1}}\xi_{\A_{1}+\A_{2}}\eta_{\A_{1}+\A_{2}} \CR
& &
\!\!\!\!\!\!\!\!\!\!
 +\G_{\A_{1}}\xi_{\A_{2}+\A_{3}}\eta_{\A_{2}+\A_{3}}
-\G_{\A_{1}}\xi_{\A_{1}+\A_{2}+\A_{3}}\eta_{\A_{1}+\A_{2}+\A_{3}}
-\xi_{\A_{1}+\A_{2}}\eta_{\A_{2}}
-\xi_{\A_{1}+\A_{2}+\A_{3}}\eta_{\A_{2}+\A_{3}}, \CR
j_{\A_{2}}(z)&=&(k+1)\pa\xi_{\A_{2}} +i \A_{+}\A_{2}\cdot\pa\vp \xi_{\A_{2}}
+\xi_{\A_{1}+\A_{2}}\B_{\A_{1}}
-\xi_{\A_{2}}\G_{\A_{3}}\B_{\A_{3}} \CR
& & -\xi_{\A_{2}}\xi_{\A_{2}+\A_{3}}\eta_{\A_{2}+\A_{3}}
+\xi_{\A_{2}+\A_{3}}\B_{\A_{3}}, \CR
J_{\A_{3}}(z)&=& -(k+2) \pa\G_{\A_{3}}-i\A_{+}\A_{3}\cdot\pa\vp
-\G_{\A_{3}}^{2}\B_{\A_{3}}
+\xi_{\A_{2}+\A_{3}}\eta_{\A_{2}}
+\xi_{\A_{1}+\A_{2}+\A_{3}}\eta_{\A_{2}+\A_{3}}, \CR
H^{i}(z)&=& -i\A_{+}\pa\vp^{i}+\A_{1}^{i}\G_{\A_{1}}\B_{\A_{1}}
+\A_{3}^{i}\G_{\A_{3}}\B_{\A_{3}} 
+\A_{2}^{i} \xi_{\A_{2}}\eta_{\A_{2}}
 +(\A_{1}+\A_{2})^{i} \xi_{\A_{1}+\A_{2}}\eta_{\A_{1}+\A_{2}} \CR
& &
+(\A_{2}+\A_{3})^{i} \xi_{\A_{2}+\A_{3}}\eta_{\A_{2}+\A_{3}}
+(\A_{1}+\A_{2}+\A_{3})^{i} \xi_{\A_{1}+\A_{2}+\A_{3}} 
\eta_{\A_{1}+\A_{2}+\A_{3}}, 
\end{eqnarray}
where $\A_{+}=\sqrt{k}$.
The currents for positive non-simple roots are obtained from those for
the simple roots. 
Introduce the operators
\begin{eqnarray}
\cL_{n}(z)&=& {1-n^{2}\over2} \A_{3}\cdot H\G^{n} -{n(n-1)\over2}
J_{-\A_{3}} \G^{n+1} + {n(n+1)\over2} J_{\A_{3}}\G^{n-1}, \CR
\cG^{1}_{n-1/2}(z)&=&-(n-1) j_{-\A_{1}-\A_{2}}\G^{n} +n
j_{-\A_{1}-\A_{2}-\A_{3}} \G^{n-1}, \CR
\cG^{2}_{n-1/2}(z)&=&-(n-1) j_{-\A_{2}}\G^{n} +n
j_{-\A_{2}-\A_{3}} \G^{n-1}, \CR
\bar{\cG}^{1}_{n-1/2}(z)&=& -(n-1) j_{\A_{1}+\A_{2}}\G^{n} - n
j_{\A_{1}+\A_{2}+\A_{3}} \G^{n-1} \CR
&& -n (n-1) \left(
\A_{3}\cdot H \G^{n-1}\tilde{\xi}_{\A_{1}+\A_{2}+\A_{3}}
+J_{-\A_{3}}\G^{n}\tilde{\xi}_{\A_{1}+\A_{2}+\A_{3}}
-J_{\A_{3}}\G^{n-2}\tilde{\xi}_{\A_{1}+\A_{2}+\A_{3}}
\right), \CR
\bar{\cG}^{2}_{n-1/2}(z)&=& -(n-1) j_{\A_{2}}\G^{n} - n
j_{\A_{2}+\A_{3}} \G^{n-1} \CR
&&
\!\!\!\!\!\!\!\!\!\!
 -n (n-1) \left(
\A_{3}\cdot H \G^{n-1}\tilde{\xi}_{\A_{2}+\A_{3}}
+J_{-\A_{3}}\G^{n}\tilde{\xi}_{\A_{2}+\A_{3}}
-J_{\A_{3}}\G^{n-2}\tilde{\xi}_{\A_{2}+\A_{3}}
\right), \CR
\cT^{+}_{n}(z)&=& J_{\A_{1}}\G^{n}+n
\tilde{\xi}_{\A_{1}+\A_{2}+\A_{3}} \tilde{\eta}_{\A_{2}}\G^{n-1}, \CR
\cT^{-}_{n}(z)&=& J_{-\A_{1}}\G^{n}+n
\tilde{\xi}_{\A_{2}+\A_{3}} \tilde{\eta}_{\A_{1}+\A_{2}}\G^{n-1}, \CR
\cT^{0}_{n}(z)&=& {1\over2}\A_{1}\cdot H\G^{n}
+{n\over2} \tilde{\xi}_{\A_{1}+\A_{2}+\A_{3}}
\tilde{\eta}_{\A_{1}+\A_{2}}\G^{n-1} 
-{n\over2} \tilde{\xi}_{\A_{2}+\A_{3}}
\tilde{\eta}_{\A_{2}}\G^{n-1} , 
\end{eqnarray}
where $\G=\G_{\A_{3}}$ and 
\begin{eqnarray}
\tilde{\xi}_{\A_{2}+\A_{3}}&=&
\xi_{\A_{2}+\A_{3}}+\G_{\A_{3}}\xi_{\A_{3}}, \CR
\tilde{\xi}_{\A_{1}+\A_{2}+\A_{3}}&=& \xi_{\A_{1}+\A_{2}+\A_{3}}
+\G_{\A_{1}} \xi_{\A_{2}+\A_{3}}+\G_{\A_{1}}\G_{\A_{3}}\xi_{\A_{2}} +
\G_{\A_{3}}\xi_{\A_{1}+\A_{2}}, \CR
\tilde{\eta}_{\A_{1}+\A_{2}}&=&\eta_{\A_{1}+\A_{2}}
-\G_{\A_{3}}\eta_{\A_{1}+\A_{2}+\A_{3}}, \CR
\tilde{\eta}_{\A_{2}}&=& \eta_{\A_{2}}-\G_{\A_{1}}\eta_{\A_{1}+\A_{2}} 
+\G_{\A_{1}} \G_{\A_{3}}\eta_{\A_{1}+\A_{2}+\A_{3}} -\G_{\A_{3}}
\eta_{\A_{2}+\A_{3}}.
\end{eqnarray}
Then the generators 
\begin{eqnarray}
L_{n}&=&\oint {d z\over2\pi i} \cL_{n}(z), \CR
G^{a}_{r}&=&\oint {dz\over 2\pi i} \cG^{a}_{r}(z), \quad 
\bar{G}^{a}_{r}=\oint {dz\over 2\pi i} \bar{\cG}^{a}_{r}(z), \CR
T^{\pm }_{n}&=&T^{1}_{n}\pm i T^{2}_{n}=\oint {d z\over2\pi i}
\cT^{\pm}_{n}(z), \quad
T^{0}_{n}=\oint {d z\over2\pi i}
\cT^{0}_{n}(z), \quad (n\in{\bf Z}, r\in {\bf Z}+{1\over2}) 
\end{eqnarray}
can be shown to satisfy  $\cN=4$ superconformal algebra with central 
charge $c=6kp$:
\begin{eqnarray}
\mbox{[} L_{m}, L_{n} \mbox{]}&=& (m-n) L_{m+n}+{c\over 12} (m^{3}-m)
\delta_{m+n,0}, \CR
\mbox{[} L_{m}, T^{i}_{n}\mbox{]}&=& -n T^{i}_{m+n}, \CR
\mbox{[} L_{m}, G^{a}_{r}\mbox{]}&=& \left( {m\over 2}-r\right)
G^{a}_{m+r}, \quad
\mbox{[} L_{m}, \bar{G}^{a}_{r}\mbox{]}= \left( {m\over 2}-r\right)
\bar{G}^{a}_{m+r}, \CR
\mbox{[} T^{i}_{m}, T^{j}_{n} \mbox{]}&=& i \varepsilon^{ijk}
T^{k}_{m+n}
+{c\over12} m \delta^{ij}\delta_{m+n,0}, \CR
\mbox{[} T^{i}_{m}, G^{a}_{r}\mbox{]}&=&
-{1\over2} (\sigma^{i})^{a}_{b} G^{b}_{m+r}, \quad
\mbox{[} T^{i}_{m}, \bar{G}^{a}_{r}\mbox{]}=
{1\over2} (\sigma^{* i})^{a}_{b} G^{b}_{m+r}, \CR
\{ G^{a}_{r}, G^{b}_{s} \}&=& 
\{ \bar{G}^{a}_{r}, \bar{G}^{b}_{s} \}=0, \CR
\{ G^{a}_{r}, \bar{G}^{b}_{s} \}&=&
\delta^{a b} L_{r+s}-(r-s) (\sigma^{i})_{a b} T^{i}_{r+s}
+{c\over 6} (4 r^{2}-1) \delta^{a b} \delta_{r+s,0},
\end{eqnarray}
where $i,j,k=1,2,3$, $a,b=1,2$, $n,m\in{\bf Z}$ and $r,s\in {\bf Z}+1/2$.

In the present work, we have constructed boundary extended superconformal
symmetry from the world-sheet affine Lie superalgebra with a
subalgebra $sl(2,R)$. 
The Wakimoto type free field realization \cite{LS}
is shown to be useful for the construction of manifestly
supersymmetric extended superconformal algebras.
We have write down only the NS sector of superconformal algebras. 
The R-sector of the algebra is obtained by the spectral flow \cite{SS}
in the
case of ${\cal N}=2,4$ extended superconformal algebras. 
For ${\cal N}=1$ case, it is necessary to find the world-sheet
current which represents the supercurrent in the R-sector.

The present construction would be generalized to other 
affine Lie superalgebras associated with $AdS_{3}$ Lie 
supergroups\cite{GuSiTo}. 
The corresponding conformal algebras would be linearly extended 
superconformal algebras.
On the other hand, the Chern-Simons theories based on general $AdS_{3}$ Lie
supergroups lead to  non-linearly extended superconformal
algebras  constructed in \cite{ESA}. 
Hence it is an interesting problem to examine the boundary
conformal field theory for general $AdS_{3}$ supergroups.
Finally, in order to relate the present construction to the type 
IIB string theory on $AdS_{3}\times S_{3}\times T^{4}$ \cite{Pe} or $K3$, 
we need to clarify the hidden affine Lie superalgebra
symmetry in the type IIB theory.
These subjects will be discussed elsewhere. 

\vskip3mm\noindent
This work is supported in part by 
the Grant-in-Aid from the Ministry of Education, Science and Culture,
Priority Area: \lq\lq Supersymmetry and Unified
Theory of Elementary Particles'' (\#707).

\newpage

\end{document}